\documentclass[letterpaper,twocolumn,english,showpacs,preprintnumbers,amsmath,amssymb,superscriptaddress,nofootinbib,pra]{revtex4}
\usepackage[T1]{fontenc}
\usepackage[latin9]{inputenc}
\usepackage{amsmath}
\usepackage{graphicx}
\usepackage{setspace}
\usepackage{amssymb}

\makeatletter

\@ifundefined{textcolor}{}
{%
 \definecolor{BLACK}{gray}{0}
 \definecolor{WHITE}{gray}{1}
 \definecolor{RED}{rgb}{1,0,0}
 \definecolor{GREEN}{rgb}{0,1,0}
 \definecolor{BLUE}{rgb}{0,0,1}
 \definecolor{CYAN}{cmyk}{1,0,0,0}
 \definecolor{MAGENTA}{cmyk}{0,1,0,0}
 \definecolor{YELLOW}{cmyk}{0,0,1,0}
 }

\makeatother

\usepackage{babel}

\begin{document}

\title{Unfrustrated Qudit Chains and their Ground States }

\author{Ramis Movassagh}

\email[Corresponding author: ]{ramis@mit.edu }

\affiliation{Department of Mathematics, Massachusetts Institute of Technology,
Cambridge, Massachusetts, U.S.A}

\author{Edward Farhi}

\affiliation{Center for Theoretical Physics, Massachusetts Institute of Technology,
Cambridge, Massachusetts, U.S.A}

\author{Jeffrey Goldstone}

\affiliation{Center for Theoretical Physics, Massachusetts Institute of Technology,
Cambridge, Massachusetts, U.S.A}

\author{Daniel Nagaj}

\affiliation{Research Center for Quantum Information, Institute of Physics, Slovak
Academy of Sciences, Bratislava, Slovakia}

\author{Tobias J. Osborne }

\affiliation{Institute for Advanced Study, Wissenschaftskolleg zu Berlin, Berlin,
Germany }

\author{Peter W. Shor}

\affiliation{Department of Mathematics, Massachusetts Institute of Technology,
Cambridge, MA}

\date{\today}


\begin{abstract}
We investigate chains of $d$ dimensional quantum spins (qudits) on
a line with generic nearest neighbor interactions without translational
invariance. We find the conditions under which these systems are not
frustrated, i.e. when the ground states are also the common ground
states of all the local terms in the Hamiltonians. The states of a
quantum spin chain are naturally represented in the Matrix Product
States (MPS) framework. Using imaginary time evolution in the MPS
ansatz, we numerically investigate the range of parameters in which
we expect the ground states to be highly entangled and find them hard
to approximate using our MPS method. 
\end{abstract}
\maketitle

\section{Introduction}

\label{sec:intro}

A system with local interactions is {\em not frustrated}, if the
global ground state of the Hamiltonian $H=\sum_{k}H_{k}$ is also
a ground state of all the local terms $H_{k}$ each of which involves
only a few particles. Frustration in a classical or quantum system
(e.g. a spin glass \cite{spinglass}) is often the reason why finding
its ground state properties is hard. A locally constrained unfrustrated
system could still have ground states that are hard to find (e.g.,
3-SAT where one needs to test whether a Boolean formula, made up of
3-literal clauses, is satisfiable by an assignment of the Boolean
variables \cite{cooklevin}).

We choose to investigate chains of $d$-dimensional quantum spins
(qudits) with 2-local nearest-neighbor interactions. Our first result
is an analytic derivation of the necessary and sufficient conditions
for such quantum systems to be unfrustrated. Second, we look at their
ground state properties and find a range of parameters where we conjecture
that these states are highly entangled and thus may be difficult to
find computationally. We then corroborate this by a numerical investigation
using a Matrix Product State (MPS) method.

The Matrix Product State description of a quantum state has proved
to be a very useful tool for the investigation of one dimensional
quantum spin chains \cite{cirac,daniel,fannes}. A pure state of a
system of $N$ interacting $d$-dimensional quantum spins can be written
in the computational basis as $|\Psi\rangle=\sum_{i_{k}=1}^{d}\psi^{i_{1}i_{2}\dots i_{N}}|i_{1}\rangle|i_{2}\rangle\dots|i_{N}\rangle$
with $d^{N}$ parameters $\psi^{\{i\}}$. For a one-dimensional chain
the coefficients $\psi^{\{i\}}$ can be conveniently expressed in
a Matrix Product State (MPS) form \cite{vidal1,frank}, \begin{eqnarray}
\psi^{i_{1}i_{2}\dots i_{N}}=\sum_{\alpha_{1},\dots,\alpha_{N-1=1}}^{\chi}\Gamma_{\alpha_{1}}^{i_{1},[1]}\Gamma_{\alpha_{1},\alpha_{2}}^{i_{2},[2]}\cdots\Gamma_{\alpha_{N-1}}^{i_{N},[N]},\label{eq:mpsdefinition}\end{eqnarray}
 providing a local description of the system in terms of matrices
$\Gamma^{i_{k},[k]}$. One arrives at this form using a series of
Schmidt decompositions \cite{chuang}. The required size of the matrices
is related to the number $\chi$ of nonzero Schmidt coefficients required
for a decomposition of the state into two subsystems. In general,
$\chi$ needs to grow like $d^{N/2}$ for the MPS to be exact. 

Is it possible to capture the essential physics of the system accurately
enough with an efficient simulation with a much smaller $\chi\thicksim\textrm{poly}(N)$,
spanning only a small part of the full Hilbert space of the system?
In our case, the qudits are arranged on a 1D lattice and only have
nearest-neighbor interactions. We could thus expect that a reduced
space might suffice for our needs. This concept is common for the
various approaches proposed for efficient (tractable on a classical
computer) numerical investigation of quantum many body systems such
as the Density Matrix Renormalization Group \cite{white}, Matrix
Product States \cite{frank}, Tensor Product States \cite{wen} and
projected entangled pair states (PEPSs) \cite{peps}.

While the MPS formulation has been shown to work very well numerically
for most one-dimensional particle systems, complexity theory issues
seem to show there must be exceptions to this rule. Finding the ground-state
energy of a one-dimensional qudit chain with $d=11$ has been shown
to be as hard as any problem in QMA \cite{q2satline,nagajthesis}.
It is not believed that classical computers can efficiently solve
problems in QMA. However, to our knowledge until now there have not
been any concrete examples (except at phase transitions) for which
MPS methods do not appear to work reasonably well. This research was
undertaken to try to discover natural examples of Hamiltonians for
which MPS cannot efficiently find or approximate the ground states.

\begin{figure}
\begin{centering}
\includegraphics[scale=0.28]{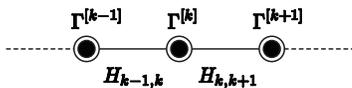} 
\par\end{centering}

\caption{A qudit chain with nearest neighbor interactions $H_{k,k+1}$ given
by \eqref{eq:hamiltonian}. The matrices $\Gamma^{[k]}$ are a local
MPS description \eqref{eq:mpsdefinition} of the state.}
\end{figure}

The paper is organized as follows. First, in Section \ref{sec:zeroenergy}
we show that the question of non-frustration for qudit chain Hamiltonians
with general nearest-neighbor interactions can be simplified to only
Hamiltonians that are sums of projector terms\cite{bravyi}. We then
analytically show under what conditions zero energy ground states
for this system exist. Second, in Section \ref{sec:mpsnumerics} we
show how to search for and approximate the ground states numerically
and analyze the efficiency of finding the required MPS. We identify
an interesting class of unfrustrated qudit chain Hamiltonians, on
which our MPS methods do not work well. Led by our numerical work,
we conjecture that these ground states are highly entangled. Finally,
we summarize our results and conclude with an outlook to further work
in Section \ref{sec:conclusions}.


\section{When is a qudit chain unfrustrated?}

\label{sec:zeroenergy}

We investigate chains of $d$-dimensional quantum particles (qudits)
with nearest-neighbor interactions. The Hamiltonian of the system,
\begin{equation}
H=\sum_{k=1}^{N-1}H_{k,k+1}\label{eq:hamiltonian}\end{equation}
 is 2-local (each $H_{k,k+1}$ acts non-trivially only on two neighboring
qudits)(Fig. 1). Our goal is to find the necessary and sufficient
conditions for the quantum system to be {\em unfrustrated} -- its
ground state is also a common ground state of all of the local terms
$H_{k,k+1}$. The local terms can be written as \begin{eqnarray}
H_{k,k+1}=E_{0}^{(k)}P_{k,k+1}^{(0)}+\sum_{p}E_{p}^{(k)}P_{k,k+1}^{(p)},\label{excited}\end{eqnarray}
 where $E_{0}^{(k)}$ is the ground state energy of $H_{k,k+1}$ and
each $P_{k,k+1}^{(p)}$ is a projector onto the subspace spanned by
the eigenstates of $H_{k,k+1}$ with energy $E_{p}^{(k)}$. The question
of existence of a common ground state of all the local terms is equivalent
to asking the same question for a Hamiltonian whose interaction terms
are \begin{eqnarray}
H_{k,k+1}' & = & \mathbb{I}_{1,\dots,k-1}\otimes P_{k,k+1}\otimes\mathbb{I}_{k+2,\dots,N},\label{eq:projham}\end{eqnarray}
 with $P_{k,k+1}=\sum_{p=1}^{r}P_{k,k+1}^{(p)}$ projecting onto the
excited states of each original interaction term $H_{k,k+1}$. When
this modified system is unfrustrated, its ground state energy is zero
(all the terms are positive semi-definite). The unfrustrated ground
state belongs to the intersection of the ground state subspaces of
each original $H_{k,k+1}$ and is annihilated by all the projector
terms.

We now choose to focus on a class of Hamiltonians for which each \begin{eqnarray}
P_{k,k+1}=\sum_{p=1}^{r}|v_{k,k+1}^{p}\rangle\langle v_{k,k+1}^{p}|\label{eq:definev}\end{eqnarray}
 is a {\em random} rank $r$ projector acting on a $d^{2}$-dimensional
Hilbert space of two qudits, chosen by picking an orthonormal set
of $r$ random vectors (a different set for every qudit pair -- we
are not assuming translational invariance). 

We now find conditions governing the existence of zero energy ground
states (from now on, called solutions in short). We do so by counting
the number of solutions possible for a subset of the chain, and then
adding another site and imposing the constraints given by the Hamiltonian.

Suppose we have a set of $s_{n}$ linearly independent solutions for
the first $n$ sites of the chain in the form \begin{equation}
\psi_{\alpha_{n}}^{i_{1},\dots,i_{n}}=\Gamma_{\alpha_{1}}^{i_{1},[1]}\Gamma_{\alpha_{1}\alpha_{2}}^{i_{2},[2]}\dots\Gamma_{\alpha_{n-2},\alpha_{n-1}}^{i_{n-1},[n-1]}\Gamma_{\alpha_{n-1},\alpha_{n}}^{i_{n},[n]}\end{equation}
 similar to MPS, with $i_{k}=1,\dots,d$ and $\alpha_{k}=1,\dots,s_{k}$;
here and below all the repeated indices are summed over. The $\Gamma$'s
satisfy the linear independence conditions%
\footnote{Note that this is not the standard MPS form, which also requires linear
independence in the other direction, i.e. \[
y_{\alpha_{k-1}}\Gamma_{\alpha_{k-1},\alpha_{k}}^{i_{k},[k]}=0,\forall i_{k},\alpha_{k}\Longleftrightarrow y_{\alpha_{k-1}}=0,\;\forall\alpha_{k-1}\]
 In this case $s_{k}$ would be the Schmidt rank for the partition
of the qudits into $(1,\dots,k)$ and $(k+1,\dots,n).$ %
} \[
\Gamma_{\alpha_{k-1},\alpha_{k}}^{i_{k},[k]}x_{\alpha_{k}}=0,\forall i_{k},\alpha_{k-1}\Longleftrightarrow x_{\alpha_{k}}=0,\;\forall\alpha_{k}.\]
 We now add one more site to the chain, impose the constraint $P_{n,n+1}$
and look for the 
zero-energy ground states for $n+1$ sites in the form \begin{equation}
\psi_{\alpha_{n+1}}^{i_{1},\dots,i_{n+1}}=\psi_{\alpha_{n-1}}^{i_{1},\dots,i_{n-1}}\Gamma_{\alpha_{n-1},\alpha_{n}}^{i_{n},[n]}\Gamma_{\alpha_{n},\alpha_{n+1}}^{i_{n+1},[n+1]}.\end{equation}
 The unknown matrix $\Gamma_{\alpha_{n},\alpha_{n+1}}^{i_{n+1},[n+1]}$
must satisfy 
\begin{eqnarray}
\langle v_{n,n+1}^{p}|i_{n}i_{n+1}\rangle\Gamma_{\alpha_{n-1},\alpha_{n}}^{i_{n},[n]}\Gamma_{\alpha_{n},\alpha_{n+1}}^{i_{n+1},[n+1]}=0\end{eqnarray}
 for all values of $\alpha_{n-1},\alpha_{n+1}$ and $p$, with $|v_{n,n+1}^{p}\rangle$
vectors defined in \eqref{eq:definev}. This results in a system of
linear equations \begin{eqnarray}
C_{p\alpha_{n-1},i_{n+1}\alpha_{n}}\Gamma_{\alpha_{n},\alpha_{n+1}}^{i_{n+1},[n+1]}=0,\end{eqnarray}
 with $C_{p\alpha_{n-1},i_{n+1}\alpha_{n}}=\langle v_{n,n+1}^{p}|i_{n}i_{n+1}\rangle\Gamma_{\alpha_{n-1},\alpha_{n}}^{i_{n},[n]}$
a matrix with dimensions $rs_{n-1}\times ds_{n}$. If $ds_{n}\geq rs_{n-1}$
and the matrix $C$ has rank $rs_{n-1}$, the conditions (9) are independent
and we can construct $ds_{n}-rs_{n-1}$ linearly independent $\Gamma_{\alpha_{n},\alpha_{n+1}}^{i_{n+1},[n+1]}$,
corresponding to solutions for the $n+1$ qudit chain (see the appendix
for further discussion of the rank of $C$). The freedom we have now
is to use only a subset of them for constructing solutions. Thus,
we obtain the formula \begin{equation}
s_{n+1}\leq ds_{n}-rs_{n-1},\label{eq:theequation}\end{equation}
valid for all $n$. The question now is how to choose $s_{n}$ 
as we go along the chain. 

The only constraint on $\Gamma_{\alpha_{1}}^{[1],i_{1}}$ is linear
independence, which requires $s_{1}\leq d$ ($s_{0}=1$ as the first
pair have $r$ constraints). 
If we choose the equality sign in the recursion relationship above
at each step we obtain $D_{n}$ linearly independent zero energy states,
where \begin{eqnarray}
D_{n}=dD_{n-1}-rD_{n-2},\end{eqnarray}
 for all $n$ with $D_{0}=1$ and $D_{1}=d$. The solution of this
recursion relation is \[
D_{n}=\frac{f^{n+1}-g^{n+1}}{f-g}\]
 with $f+g=d$ and $fg=r$. Hence, \[
f=\frac{d}{2}+\sqrt{\frac{d^{2}}{4}-r},\quad g=\frac{d}{2}-\sqrt{\frac{d^{2}}{4}-r}.\]
 There are three interesting regimes for $r$ and $d$ which yield
different behaviors of $D_{n}$(Figure 2): 
\begin{enumerate}
\item $r>\frac{d^{2}}{4}$ gives $D_{n}=r^{\frac{n}{2}}\frac{\sin(n+1)\theta}{\sin\theta}$
with $\cos\theta=\frac{d}{2\sqrt{r}}$. $D_{n}$ becomes negative
when $n+1>\frac{\pi}{\theta}$ and thus no zero energy states can
be constructed for a long chain if $r>\frac{d^{2}}{4}$. 
\item $r=\frac{d^{2}}{4}$ results in $D_{n}=\left(\frac{d}{2}\right)^{n}(n+1)$,
an exponential growth in $n$ (except when $d=2$, which gives linear
growth).

\item $r<\frac{d^{2}}{4}$ implies $f>\frac{d}{2}$ and $f>g$ so for large
$n$, $D_{n}\sim f^{n}\left(1-\frac{g}{f}\right)^{-1}$ and the number
of zero energy states grows exponentially. 
\end{enumerate}
Any set of $s_{n}$ that satisfies the inequality \eqref{eq:theequation}
must have $s_{n}\leqslant D_{n}$. To show this, we rewrite \eqref{eq:theequation}
as 

$s_{0}=1$

$s_{1}-ds_{0}=-u_{1}$

$s_{n}-ds_{n-1}+rs_{n-2}=-u_{n},\; n\geqslant2$

with $u_{n}\geqslant0,\; n\geqslant1$.

\begin{flushleft}
These relations can be inverted to give
\par\end{flushleft}

$s_{n}=D_{n}-\sum_{l=1}^{n}u_{l}D_{n-l},\; n\geqslant1$

\begin{flushleft}
from which $s_{n}\leqslant D_{n}$ follows at once.
\par\end{flushleft}

This means that in case 1, it is still not possible to construct solutions
for a long chain. In cases 2 and 3, we can construct sets of states
with $s_{n}$ growing more slowly than $D_{n}$.
\begin{itemize}
\item For example when $d\leq r\leq\frac{d^{2}}{4}$, the recursion relation
\eqref{eq:theequation} can be satisfied also by $s_{n}=h^{n}$ provided
that $h^{2}-dh+r\leq0$. This requires $g\leq h\leq f$, so the lower
bound on $h$ is the least integer $\geq g$. 
\item On the other hand, for $r<d$ the simple choice $s_{n}=1$ also satisfies
the recursion \eqref{eq:theequation}. This means one can just solve
the system from left to right as a linear system of equations. This
results in a product state solution in the form \begin{equation}
\psi^{i_{1}i_{2}\dots i_{n}}=\psi^{[1],i_{1}}\psi^{[2],i_{2}}\cdots\psi^{[n],i_{n}}\end{equation}
 which we can construct by starting with any $\psi^{[1]}$ and finding
every $\psi^{[n+1]}$ from the previous ones. 
\end{itemize}
\begin{figure}
\includegraphics[width=5cm]{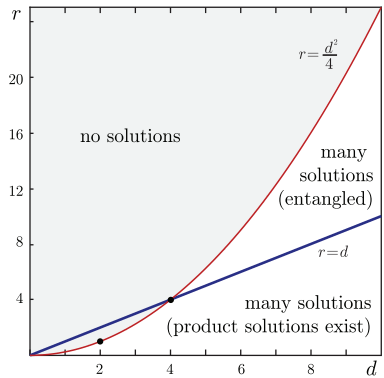}

\caption{The existence of zero energy ground states for a qudit chain with
$d$-dimensional qudits and $r$ projectors per pair. We highlight
two notable cases: $d=2,r=1$ and $d=4,r=4$.}

\label{fig:parameters} %
\end{figure}


\section{Numerical investigation using Matrix Product States}

\label{sec:mpsnumerics}

\begin{figure}
\begin{centering}
\includegraphics[width=8cm]{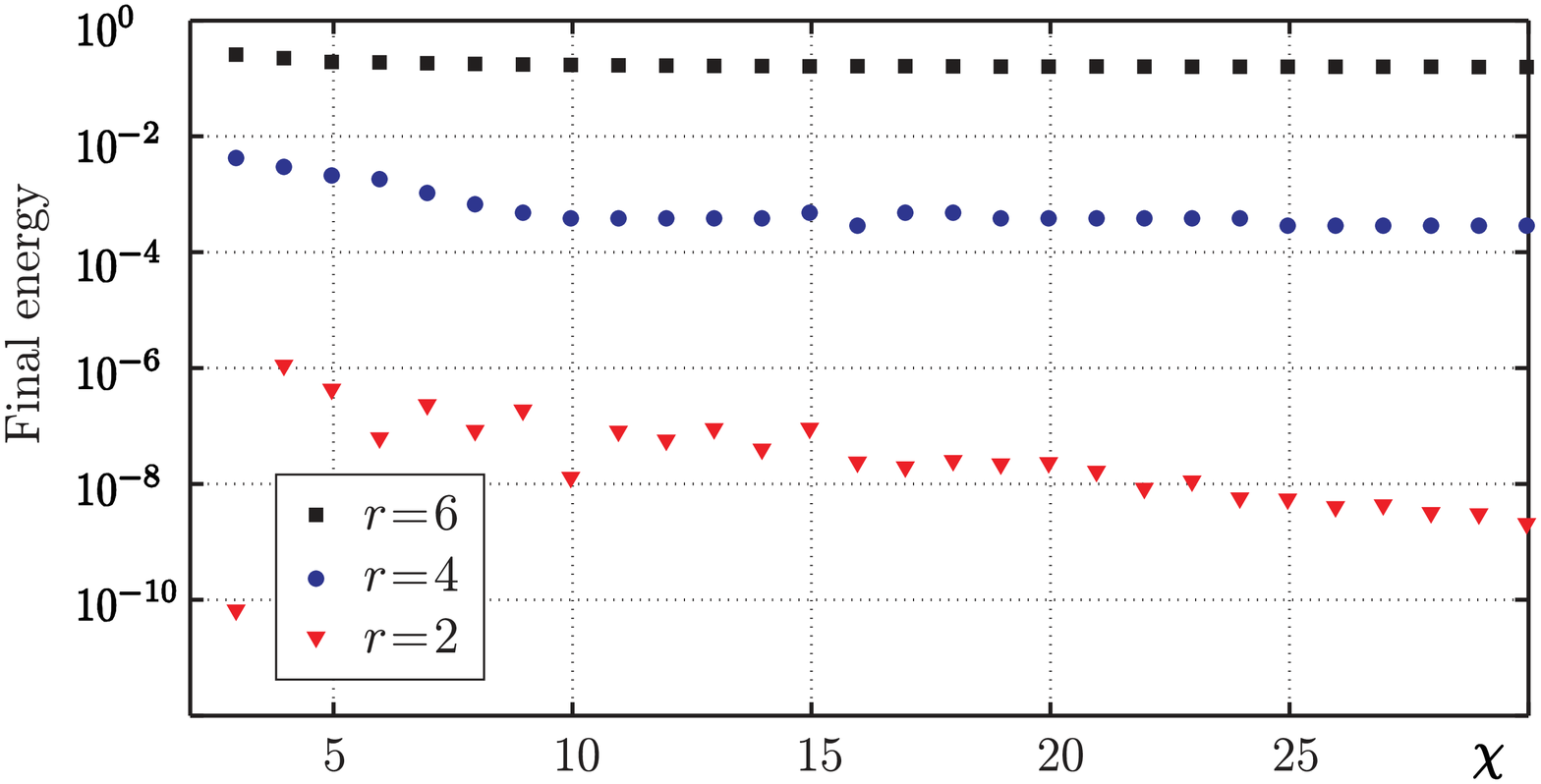} 
\par\end{centering}

\caption{(Color online) Ground state energy from imaginary time evolution vs.
$\chi$ for different ranks of the Hamiltonian. This is a plot for
d = 4, and projector ranks of 2, 4, 6. Exact description would require
$\chi=d^{\frac{N}{2}}=2^{20}.$}
\end{figure}

In this Section we numerically search for the ground states of our
class of random projector Hamiltonians \eqref{eq:projham}. We probe
the relations obtained in the previous Section, and see how well the
energy coming from our small-$\chi$ MPS imaginary time evolution
converges to zero. The numerical technique we use is similar to Vidal's
\cite{vidal1,vidal2}. We use imaginary time evolution to bring the
system from a known state to its ground state: $|\Psi_{\textrm{grd}}\rangle=\lim_{\tau\rightarrow\infty}\frac{e^{-H\tau}|\Psi_{0}\rangle}{||e^{-H\tau}|\Psi_{0}\rangle||}$.
In our numerical work we normalize the state after every time step
\cite{daniel}. We start from a uniform superposition of all the states
and Trotterize by evolving first the odd pairs of sites and then the
even pairs. Our experimentation with the parameters for a linear chain
of length $N=20$ is shown in Figures 3 , 4 and 5; all the plots are
on semi-log scale and the quantities being plotted are dimensionless.

We see that for $r<d$ the final energy converges to the zero energy
ground state relatively fast with $\chi\ll d^{N/2}$. This can be
seen in all the figures by the lowest curves (marked by triangles).
As can be seen the final energy obtained from imaginary time evolution
tends toward zero with a steep slope, indicating that the ground state
can be approximated efficiently with a small $\chi$ in MPS ansatz. 

\begin{figure}
\begin{centering}
\includegraphics[width=8cm]{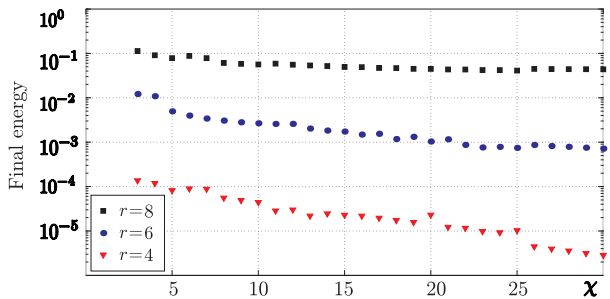} 
\par\end{centering}

\caption{(Color online) This is a plot for d = 5, and projector ranks of 4,
6, 8. Exact description would require $\chi=5^{10}$.}
\end{figure}

\begin{figure}
\begin{centering}
\includegraphics[width=8cm]{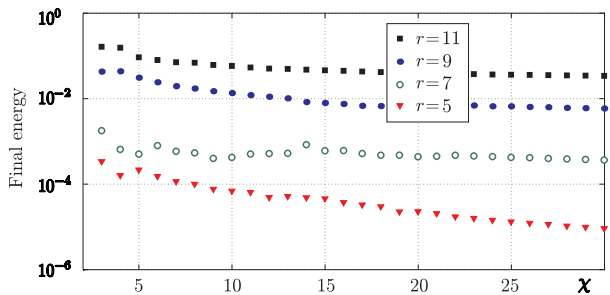}
\par\end{centering}

\caption{(Color online) The case of d = 6, and projector ranks: 5, 7, 9, 11.
Exact description in general would require $\chi=6^{10}$.}
\end{figure}

The \textit{$r>d^{2}/4$ }case, marked by squares, is shown by the
top curves in all the figures. One sees that the final energy plateaus
relatively fast in all three cases. This shows that the numerics have
converged to a nonzero value and that increasing $\chi$ will not
yield a lower value of energy. Therefore, the numerical results suggest
that there are no ground states with zero energy. 

In the previous section we analytically showed that when $d\leq r\leq d^{2}/4$
there are many zero energy ground states. However, when we try to
numerically find these states we see that the final energy converges
to zero slowly. This is shown in all the Figures by the curves marked
by circles. Out of these there are the critical cases, where $r=\frac{d^{2}}{4}$.
These correspond to the curves marked by closed circles in Figures
3 and 5. The numerical investigation of the case $d\leq r\leq d^{2}/4$
is interesting because it suggests that for large number of spins
finding the ground state with small $\chi$, tractable on a normal
computer, is very hard. We interpret this as high amount of entanglement
among the zero energy ground states and leave the analytical proof
of this statement for a follow up paper.\\


\section{Conclusions}

\label{sec:conclusions}

We have investigated the no-frustration conditions for a system of
qudits on a line with $d$ states per site and random rank $r$ local
projector Hamiltonians acting between the nearest neighbor sites.
We proved that there are no ground states with zero energy for $r>\frac{d^{2}}{4}$
and sufficiently large $N$. The system is not frustrated for $r\leq\frac{d^{2}}{4}$.
This second parameter region further splits into two. For $d\leq r\leq\frac{d^{2}}{4}$,
many entangled zero energy ground states exist. On the other hand,
for $r<d$ we can also construct separable zero-energy ground states
(see also Figure \ref{fig:parameters}).

We have verified the above numerically, in particular we have seen
that when $d\leq r\leq d^{2}/4$ approximating the ground state energy
(finding the ground states) is hard as the states seem to be highly
entangled. Future work entails the investigation of the energy gap
\cite{hastings} and the amount of entanglement in the system as a
function of the parameters of the chain. Furthermore, we would like
to address how far from an eigenstate of the Hamiltonian is the wave
function after the truncations are made (i.e. as a function of $\chi$).
Finally, we would like to quantify the nature of the convergence to
the ground state starting from an arbitrary state in this framework.


\section{Acknowledgments}

\label{sec:acknowledgements}

RM would like to thank Salman Beigi, Michael Artin, Sam Gutmann and
especially Alan Edelman for fruitful discussions. RM and PWS would
like to thank the National Science Foundation for the support through
grant number CCF-0829421. EF and JG were supported in part by funds
provided by the U.S. Department of Energy under cooperative research
agreement DE-FG02-94ER40818, the W. M. Keck Foundation Center for
Extreme Quantum Information Theory, the U.S. Army Research Laboratory's
Army Research Office through grant number W911NF-09-1-0438, the National
Science Foundation through grant number CCF-0829421. DN thanks Eddie
Farhi's group for their hospitality and gratefully acknowledges support
from the Slovak Research and Development Agency under the contract
No. APVV LPP-0430-09 and from the European project OP CE QUTE ITMS
NFP 26240120009.




\section{Appendix}

$\;$

We now address the question of the rank of the matrix $C$ in equation
(9). This $C$, which gives the constraints on $\Gamma^{[n+1]}$ depends
on $|v_{n,n+1}^{p}\rangle$, and through $\Gamma^{[n]}$ on all the
$|v_{k,k+1}^{p}\rangle$ with $k<n$. We would like to say that for
random $|v\rangle$ and when $s_{k}<D_{k}$ for random choices of
the $s_{k}$ dimensional subspace, the rank of $C$ is generically
the maximum rank allowed, $\min(rs_{n-1},ds_{n})$. This is not obviously
true. In particular for the case $s_{k}=D_{k}$ in the regime $r\leq\frac{d^{2}}{4}$,
to which we restrict ourselves from now on, $D_{n}$ grows exponentially
in $n$, while the number of parameters in the $|v_{k,k+1}^{p}\rangle$
on which $C$ depends only grows linearly. Thus $C$ is far from a
generic matrix of its size, but we now prove that its rank is indeed
$rD_{n-1}.$

The argument used by Laumann et al \cite{laumann} to prove their
`geometrization theorem' also applies to our problem. It shows that
for a chain of N qudits with random $|v_{k,k+1}^{p}\rangle$, i.e.
for a Hamiltonian $H$ as in equations (2), (4) and (5) the number
of zero-energy states, i.e. $\dim(\ker(H))$, is with probability
one (which is what we mean by generic) equal to its minimum value.
The calculation in section II leading to the recursion relation (11)
and its solution, shows that in the regime $r\leq d^{2}/4$ this minimum
is $\geq D_{N},$ since if the rank of the $rD_{k-1}\times dD_{k}$
matrix $C$ is ever less than $rD_{k-1}$ we can choose $s_{k+1}=D_{k+1}$.
Hence it is sufficient to find a single set of $|v_{k,k+1}^{p}\rangle$
for which $\dim(\ker(H))=D_{N}$ to prove that $D_{N}$ is the generic
value, i.e. that greater values occur with probability zero. This
implies that the rank of each $C$ is generically $rD_{k-1}$, since
otherwise at the first $k$ where $C$ had smaller rank we could construct
more than $D_{k+1}$ solutions for a chain of length $k+1$. 

\begin{singlespace}
We construct $|v_{k,k+1}^{p}\rangle$ with the property $\langle v_{k,k+1}^{p}|i_{k}i_{k+1}\rangle=0$
unless $i_{k}\leq\frac{d}{2}$ and $i_{k+1}>\frac{d}{2}$. This can
be done for $r$ linearly independent $|v^{p}\rangle$ if $r\leq\frac{d^{2}}{4}$.
We now assume $d$ is even; the modifications for $d$ odd are obvious.
We proceed by induction on $n$. Assume that in each $\Gamma_{\alpha_{k-1},\alpha_{k}}^{i_{k},[k]}$
with $k\leq n$, $\alpha_{k}$ runs from $1$ to $D_{k}$. From the
definition of $C$ (following equation 9) and the special choice of
$|v\rangle$, $C_{p\alpha_{n-1},i_{n+1}\alpha_{n}}=0$ for $i_{n+1}\leq\frac{d}{2}$
and so from equation (9) $\Gamma_{\alpha_{n},\alpha_{n+1}}^{i_{n+1},[n+1]}$
is unconstrained for $i_{n+1}\leq\frac{d}{2}$. This allows us to
choose, for $1\leq\alpha_{n+1}\leq\frac{d}{2}D_{n}$,
\end{singlespace}

\noindent $\quad\Gamma_{\alpha_{n},\alpha_{n+1}}^{i_{n+1},[n+1]}=1$
when $\alpha_{n+1}=\frac{d}{2}(\alpha_{n}-1)+i_{n+1}$ 

\noindent with $1\leq\alpha_{n}\leq D_{n}$, $1\leq i_{n+1}\leq\frac{d}{2}$

\noindent $\quad\Gamma_{\alpha_{n},\alpha_{n+1}}^{i_{n+1},[n+1]}=0$
otherwise.

\begin{singlespace}
As part of our induction, we assume that for $1\leq\alpha_{n}\leq\frac{d}{2}D_{n-1}$,
\end{singlespace}

\noindent $\quad\Gamma_{\alpha_{n-1},\alpha_{n}}^{i_{n},[n]}=1$ when
$\alpha_{n}=\frac{d}{2}(\alpha_{n-1}-1)+i_{n}$ 

\noindent with $1\leq\alpha_{n-1}\leq D_{n-1}$, $1\leq i_{n}\leq\frac{d}{2}$ 

\noindent $\quad\Gamma_{\alpha_{n-1},\alpha_{n}}^{i_{n},[n]}=0$ otherwise.

Now we can show that the rows of $C_{p\alpha_{n-1},i_{n+1}\alpha_{n}}$
are linearly independent. For if, $\sum_{p,\alpha_{n-1}}y_{p,\alpha_{n-1}}C_{p\alpha_{n-1},i_{n+1}\alpha_{n}}=0$
for all $i_{n+1},\alpha_{n}$ this is true in particular for all $i_{n+1}>d/2$,
$\alpha_{n}\leq\frac{d}{2}D_{n-1}$, when it becomes $\sum_{p}y_{p\alpha_{n-1}}\langle v_{n,n+1}^{p}|i_{n}i_{n+1}\rangle=0$
for all $i_{n}\leq\frac{d}{2}$, $i_{n+1}>\frac{d}{2}$, and $\alpha_{n-1}\leq D_{n-1}$.
Since the $|v^{p}\rangle$ are linearly independent, this is only
true if $y_{p\alpha_{n-1}}=0$ for all $p$, $\alpha_{n-1}$. Hence
the rank of $C$ is $rD_{n-1}$ and $\alpha_{n+1}$ can take altogether
$dD_{n}-rD_{n-1}=D_{n+1}$ values, which is what we wanted to prove.
\end{document}